\documentclass[aps,prb,twocolumn,groupedaddress]{revtex4}
\usepackage{graphicx}
\begin{document}

\title{Scaling of the conductance distribution near
the Anderson transition}

\author{Keith Slevin}
\affiliation{Department of Physics, Graduate School of Science,
Osaka University, \\ 1-1 Machikaneyama, Toyonaka,
Osaka 560-0043, Japan.}

\author{Peter Marko\v{s}}
\affiliation{Institute of Physics, Slovak Academy of Sciences,
D\'ubravsk\'a cesta 9, 842 28 Bratislava, Slovakia.}

\author{Tomi Ohtsuki}
\affiliation{Department of Physics, Sophia University,
Kioi-cho 7-1, Chiyoda-ku, Tokyo 102-8554, Japan.}

\date{\today}

\begin{abstract}
The scaling hypothesis is the foundation of our understanding 
of the Anderson transition.  
It has long been realised that the conductance of a disordered system 
is a fluctuating quantity and often assumed, but never demonstrated
in the literature, that the conductance distribution obeys a single 
parameter scaling law.
We present a clear cut numerical demonstration of the scaling 
of the conductance distribution in the critical regime.
\end{abstract}


\maketitle

\section{Introduction}

The single parameter scaling hypothesis 
of Abrahams {\it et al.} is the basis of our understanding 
of the Anderson metal-insulator transition in disordered 
systems.\cite{abrahams:79}
In Ref. \onlinecite{abrahams:79} it was proposed that the 
zero temperature conductance $G=(e^2/h)g$,
as measured by the ^^ ^^ Thouless number", obeys a single 
parameter scaling law.
However, the large sample to sample fluctuations in the conductance
of disordered systems were not explicitly considered.
In the critical and localised regimes the fluctuations
are of the same order as the mean conductance.
(The relation of the mean conductance to the Thouless number
is discussed in Ref. \onlinecite{braun:97}.)
This led to suggestions that the 
scaling hypothesis should be reformulated in terms of the typical 
conductance,\cite{anderson:80} or perhaps the distribution of 
conductance.\cite{shapiro:86, shapiro:87}

For a disordered system of size $L$ in $d=2+\epsilon$ dimensions,
Altshuler {\it et al.}  estimated the 
cumulants $c_n$ of the conductance distribution using a field
theoretic method.\cite{altshuler:89}
At the mobility edge they found that
\begin{equation}
c_n \left( L \right) = \left\{
\begin{array}{l}
\epsilon^{n-2},\  n\le n_0 \approx 1/\epsilon \\
\left( L/\ell \right)^{\epsilon n^2 - 2 n}, \ n>n_0
\end{array}
\right. 
\label{cn}
\end{equation}
If single parameter scaling holds, the only relevant 
length should be the correlation length $\xi$, and the appearance of the mean 
free path $\ell$ in the expression for the higher cumulants is unexpected.
(In the insulating regime $\xi$ is the localisation 
length, while in the metallic regime it is the correlation length.)
Shapiro reconstructed the critical conductance distribution 
from (\ref{cn}) and showed that, appearances to the contrary,
these cumulants are consistent with a single
parameter scaling of the distribution.\cite{shapiro:90}

To apply results for $d=2+\epsilon$ to three dimensions we must   
make a questionable extrapolation to $\epsilon=1$. 
While the Anderson transition occurs at weak disorder $k_F\ell\gg1$ 
when $\epsilon \ll 1$, it occurs at strong disorder $k_F \ell \approx 1$ in
three dimensions. (Here  $k_F$ is the Fermi wave number.) 
Comparison of the distribution obtained by Shapiro 
with numerical results  shows
that the behaviour of the conductance distribution at large $g$, and also the
non-universal behaviour of the higher cumulants in (\ref{cn}), is
qualitatively incorrect.\cite{slevin:97,markos:99}

To try to overcome the limitation to weak disorder Cohen 
{\it et al.} 
used a Migdal-Kadanoff type real space renormalization scheme.\cite{cohen:88} 
They found that the scaling of the conductance 
distribution is described by two parameters;
only in the limit of weak disorder is single parameter scaling recovered.
However, and as pointed out by Cohen {\it et al.}, the Migdal-Kadanoff 
scheme involves an uncontrolled approximation and some of the results
obtained with it are known not to be correct.
For example, in the metallic regime the predicted conductance fluctuations 
are too large and in disagreement with the theoretically well established 
and experimentally verified phenomena of universal conductance 
fluctuations.\cite{altshuler:85,lee:85}
They concluded that while the Migdal-Kadanoff scheme may be 
exact for hierarchical lattices, in three dimensions it is primarily of 
pedagogical value.\cite{cohen:95}

Recent work on Anderson localisation in one dimension has 
highlighted the importance of a second length scale $l_s$.\cite{deych:00}
Deych {\it et al.} demonstrated the existence of a crossover between 
single parameter and two parameter scaling regimes dependent on the 
ratio of $l_s$ to the localisation length $\xi$.
Single parameter scaling is observed when $\xi > l_s$, and two parameter 
scaling when $\xi < l_s$. The implications of this
result for Anderson localisation in higher dimensions are not yet clear.

Numerical studies of the Anderson model have demonstrated single 
parameter scaling 
of the localisation length of electrons in quasi-one dimensional
systems, \cite{slevin:99} and 
also of the mean resistance, mean conductance and typical
conductance near the Anderson transition in three dimensions.\cite{slevin:01}
Of themselves these studies do not rule out a two parameter scaling of
the conductance distribution.
For example, in the Migdal-Kadanoff scheme the mean
of the logarithm of resistance obeys a single parameter scaling 
law while at the same
time the distribution obeys a two parameter scaling law.
However, numerical studies have also demonstrated the 
existence of a universal size independent critical distribution that 
is accessible by varying only a single parameter.\cite{slevin:97} 
This observation makes a two parameter scaling unlikely; two parameter scaling
should require that both parameters be varied simultaneously
to access the critical point. 
Nevertheless, given that the scaling hypothesis is the foundation upon which 
our understanding of transport in disordered systems rests,
we feel that a clear cut demonstration of the scaling
of the distribution of conductance, free of questionable
extrapolations or uncontrolled approximations, is called for.

\section{Method}

Following Ref. \onlinecite{shapiro:87} a single parameter scaling
law for the conductance distribution $p_L(g)$ of a
three dimensional system of linear dimension $L$
can be mathematically formulated as follows 
\begin{equation}
p_L(g) \simeq F(g;X),
\label{SPSa}
\end{equation}
where $X$ is a parameter which must obey a single parameter scaling law
\begin{equation}
\frac{{\mathrm d} \ln X}{{\mathrm d} \ln L} = \beta(X).
\label{SPSb}
\end{equation}
A limiting process is implicit in (\ref{SPSa}); we refer the 
reader to Ref. \onlinecite{shapiro:87} for a detailed discussion.
The parameter $X$ need not be one of the moments of the distribution.

At first sight it appears that we must know the functional form of the
function $F$ in (\ref{SPSa}) in order to verify single parameter
scaling of the distribution numerically.
In fact, this is not so.
The procedure which we have adopted is to analyse the scaling of the
percentiles of the distribution.
The precise definition of the percentile $g_q$ is
\begin{equation}
q=\int_{0}^{g_q} p_L(g) {\mathrm d} g
\end{equation}
where $0 \le q\le 1$. 
By establishing single parameter scaling for a representative set of 
percentiles we indirectly establish (\ref{SPSa}) and (\ref{SPSb}),
{\it provided} that the scaling of different percentiles are consistent.
When considering the percentiles it is not necessary to distinguish $g$,
$\ln g$ or $1/g$ as it is when considering average quantities.

We have analysed the conductance distribution 
of the Anderson model numerically.
The motion of the electrons is described by 
\begin{equation}
 H = V \sum_{<i,j>} C_i^{\dagger}C_j +
     \sum_i W_i C_i^{\dagger}C_i ,
\label{Hamiltonian}
\end{equation}
where $C_i^{\dagger}(C_i)$ is the creation (annihilation)
operator of an electron at the site $i$ of a three
dimensional cubic lattice.
The amplitude of the random potential at site $i$ is $W_i$.
Hopping is restricted to nearest neighbours and its
amplitude was taken as the unit of energy, $V=1$.
We assumed a box distribution with each $W_i$ uniformly distributed 
on the interval $[-W/2,W/2]$.
In what follows we refer to the  strength of the potential fluctuations 
$W$ as the disorder.
The numerical method used is described in Ref. \onlinecite{pendry:92}.
The two terminal
zero temperature conductance was evaluated using the Landauer 
formula 
\begin{equation}
g=2\mathrm{tr} t^{\dag}t
\end{equation}
where $t$ is the transmission matrix
describing the propagation of electrons from one contact to the 
other.\cite{economou:81,fisher:81}

The conductance distribution depends on the system size $L$, the 
disorder $W$, the Fermi energy $E_F$ and
the boundary conditions. We set $E_F=0.5$ and imposed fixed
boundary conditions in the transverse directions. 
We accumulated data for the disorder range $15\le W\le 18$ and 
system sizes $6\le L\le 18$. At the extremes of the disorder range the
localisation (correlation) length is of the same order as the 
system size,\cite{mackinnon:83}
so that our data covers the critical regime.

To estimate $g_q$ we simulated 1,000,000 realisations
of the random potential and calculated the conductance for each realisation.
(For $L=18$ the number of realisations was approximately
500,000.)
We sorted the data into ascending order and our estimate of $g_q$ is then
the $n=\left[qN_d\right]$th datum in this list where $\left[x\right]$ is the integer
part of $x$.
When fitting the numerical data it is also necessary to have an
estimate of the accuracy of the percentiles. 
Following the standard method we used the binomial distribution 
to estimate the likely accuracy of the percentile. We define
\begin{equation}
\Delta n=\sqrt{N_dq\left(1-q \right)},
\end{equation}
and look up the 
$\left(n+\Delta n \right)$th and
$\left(n-\Delta n + 1\right)$th data in the list and calculate the differences with $g_q$.
Our estimate of the accuracy is then the largest of these two differences.
In practice, we found that the accuracy of all
the percentiles were comparable, being of the order of $0.2\%$.
The data were then fitted with the finite size scaling forms below by minimizing
the $\chi^2$ statistic in the usual way.

To fit the system size and disorder dependence of the percentile we supposed
a single parameter scaling law but allowed for deviations from scaling due to
an irrelevant scaling variable and non-linearity of the scaling 
variables.\cite{slevin:99}
We fitted the data to
\begin{equation}
\ln g_q = F \left(\psi, \phi \right),
\label{fit}
\end{equation}
where $\psi$ is the relevant scaling variable and $\phi$ is the irrelevant 
scaling variable.
We approximated this scaling function by its first order expansion
in the irrelevant scaling variable
\begin{equation}
\ln g_q = F_0 \left(\psi \right) + \phi F_1 \left(\psi \right).
\end{equation}
We expanded each scaling function as a power series
\begin{equation}
F_0\left(x \right) = \ln \left(g_q \right)_c + x + a_2 x^2 + \cdots
+ a_{n_0}x^{n_0}
\end{equation}
\begin{equation}
F_1\left(x \right) = 1 + b_1 x + b_2 x^2 + \cdots
+ b_{n_1}x^{n_1}
\end{equation}
Here $\left(g_{q} \right)_c$ is the critical value of the percentile.
The scaling variables were approximated by expansions in terms of the 
dimensionless disorder 
\begin{equation}
w=(W_c-W)/W_c,
\end{equation}
where $W_c$ is the critical
disorder separating the insulating and metallic phases,
\begin{equation}
\psi = L^{1/\nu} \left(\psi_1 w + \psi_2 w^2 + \cdots + \psi_{n_{\psi}} w^{n_{\psi}} \right)
\end{equation}
and
\begin{equation}
\phi = L^{y} \left(\phi_0 + \phi_1 w + \phi_2 w^2 + \cdots + \phi_{n_{\phi}} w^{n_{\phi}} \right). 
\end{equation}
The critical exponent $\nu$ describes the divergence of the 
localisation (correlation) length 
as the transition is approached. 
\begin{equation}
\xi = \xi_{\pm} \left| \psi_1 w + \psi_2 w^2 + \cdots + \psi_{n_{\psi}} w^{n_{\psi}} \right|^{-\nu}
\end{equation}
The constants $\xi_{\pm}$, and hence the absolute scale of the localisation 
(correlation) length 
$\xi$, cannot be determined from the fit.
The decay of the irrelevant scaling variable with system size is described by the
exponent $y<0$.
Redundancy in the definition of the fitting parameters between the 
coefficients in the 
expansions of $F_0$ and $F_1$ and the expansions 
of $\psi$ and $\phi$ are eliminated  
by setting some of the expansion coefficients of $F_0$ 
and $F_1$ to unity as shown.
This choice is also necessary if $F_0$ and $F_1$ are to be universal.
The total number of parameters is $N_p = n_0 + n_1 + n_{\psi} + n_{\phi}+4$.

The minimum of $\chi^2$ was
found using the DRNLIN routine of the IMSL numerical library. The starting values 
of the fitting parameters supplied to DRNLIN are in the region 
$\nu\approx 1.6$, $W_c\approx 16.5$, $y\approx -3$, $\psi_1\approx 1$. We set
the initial value of $\ln \left( g_q \right)_c$ to a value close to its best fit 
value by visual inspection of the raw data and all other parameters were initially zero.
The results of the fitting procedure are not especially sensitive to the
choice of the starting values.
A number of fits corresponding to different choices of $n_0$, $n_1$, $n_{\psi}$ and
$n_{\phi}$ are possible and a selection criterion is necessary.
We set a cut off for the goodness of fit probability $Q$ at $Q=0.1$ and searched
for the fit which requires the fewest parameters which satisfies this.
Broadly speaking all sensible choices of $n_0$, $n_1$, $n_{\psi}$ and
$n_{\phi}$ lead to consistent estimates of the critical parameters.
The goodness of fit and the accuracy of the fitted parameters were estimated
using Monte Carlo simulations of synthetic data sets.\cite{NumRep}

\section{Results}

Results for the $q=0.025$, $0.17$, $0.5$ and $0.83$ percentiles are shown
in Table \ref{T1}.
(For a normal distribution with mean $\mu$ and variance $\sigma^2$, these choices
would correspond to the points $\mu-2\sigma, \mu-\sigma, \mu, \mu+\sigma$
in the distribution.) 
Precise details of the fits are given in Table \ref{T2}.
The estimates of the irrelevant exponent are consistent with those obtained
in Ref. \onlinecite{slevin:01} and are not shown again here.

Data for the $q=0.17$ percentile are plotted in Figure 1. In Figure 2 the 
same data, after subtraction of corrections to scaling, are re-plotted to 
demonstrate single parameter scaling.
This is done by plotting the corrected data as a function of the ratio of the
systems size $L$ to the localisation (correlation) length $\xi$.
When displayed in this way the data fall on two different curves corresponding 
to the localised (lower curve) and the delocalised (upper curve) regimes. 
The two curves are described by two scaling functions $F_+$ and $F_-$ which are 
derived from $F_0$,
\begin{equation}
F_+\left( x \right) = \ln \left(g_q \right)_c + x^{1/\nu} +   \cdots +
a_{n_0}x^{n_0/\nu},
\label{fplus}
\end{equation}
\begin{equation}
F_-\left(x \right) = \ln \left(g_q \right)_c - x^{1/\nu} +   \cdots +
\left(-1 \right)^{n_0} a_{n_0}x^{n_0/\nu}.
\label{fminus}
\end{equation}
Data on the metallic branch follow
\begin{equation}
\ln g_q = F_+\left( \frac{L}{\xi_+} \right)
\end{equation}
while data on the the insulating branch follow
\begin{equation}
\ln g_q = F_-\left( \frac{L}{\xi_-} \right)
\end{equation}
For completeness some representative data for the median $(q=0.5)$ 
and $q=0.87$ percentiles appear in Figures 3 and 4 respectively.

\begin{figure}[ht]
\includegraphics[width=1.0\linewidth]{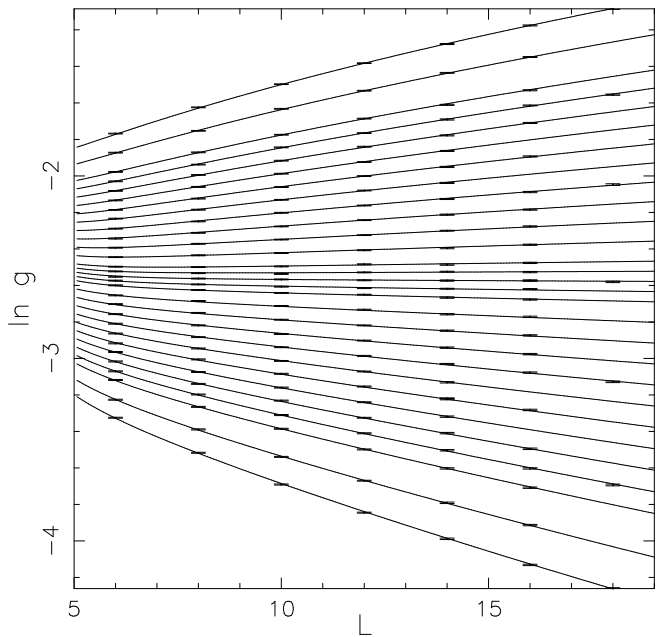}
\caption{\label{}The $q=0.17$ percentile of conductance distribution for a 
disordered $L\times L\times L$ systems
versus system size $L$ for disorder $W$ in the range $[15,18]$. The lines
are the fit of Eq. \ref{fit}.}
\end{figure}

We also analysed the  $q=0.975$ percentile of the distribution but were
unable to convincingly fit its systems size and disorder dependence.
The origin of the difficulties may be the large corrections to 
scaling encountered for this percentile.
Larger systems sizes will probably be needed for a definitive analysis
of the high conductance tail of the distribution.

\begin{figure}[ht]
\includegraphics[width=1.0\linewidth]{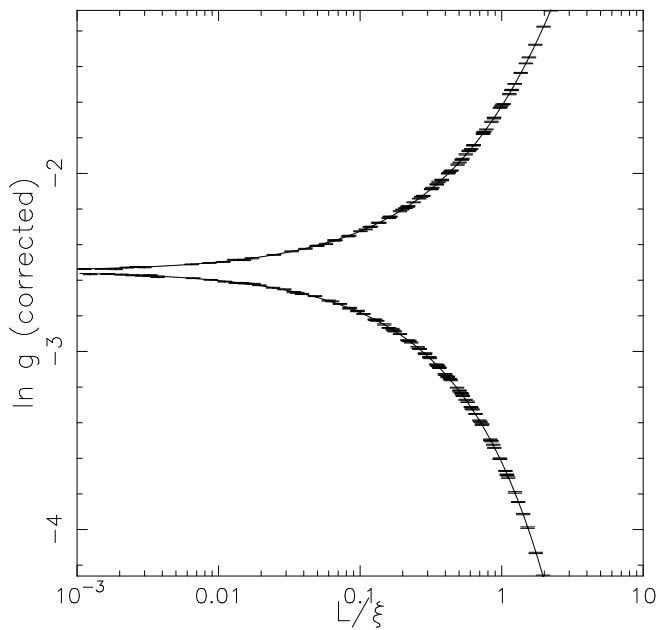}
\caption{The data of Figure 1, after subtraction of corrections to scaling,
re-plotted as function of the ratio $L/\xi$ to display single parameter scaling.
The lines are the scaling functions (\ref{fplus}) and 
(\ref{fminus}) described in the text.}
\end{figure} 

\begin{table}
\caption{\label{T1}The estimated critical values for each percentile
and 95\% confidence intervals.}
\begin{ruledtabular}
\begin{tabular}{|l|lll|}
$q$ & $W_c$           &  $\ln \left(g_{q} \right)_c$           & $\nu$           \\ \hline
0.025      & $16.48\pm.02$   &  $-4.14\pm.03$   & $1.56\pm.03$    \\
0.17       & $16.47\pm.01$   &  $-2.55\pm.01$   & $1.56\pm.01$    \\
0.5        & $16.48\pm.04$   &  $-1.08\pm.03$   & $1.59\pm.03$    \\
0.83       & $16.46\pm.05$   &  $0.13\pm.03$   & $1.60\pm.04$    \\
\end{tabular}
\end{ruledtabular}
\end{table}

\begin{table}
\caption{\label{T2}The details of the fit. $N_d$ is the number of data.}
\begin{ruledtabular}
\begin{tabular}{|l|lllll|l|l|l|}
$q$ & $n_0$ & $n_{\psi}$ & $n_1$ & $n_{\phi}$ & $N_p$ & $N_d$ & $\chi^2$ & $Q$ \\ \hline
0.025    & 3   &    2         &  1    &  0         & 10    & 179   &  159     & 0.7 \\
0.17     & 2   &    2         &  1    &  0         & 9     & 179   &  155     & 0.8 \\
0.5      & 3   &    2         &  1    &  0         & 10    & 179   &  172     & 0.4 \\
0.83     & 3   &    2         &  1    &  0         & 10    & 179   &  180     & 0.3 \\
\end{tabular}
\end{ruledtabular}
\end{table}

\begin{figure}[ht]
\includegraphics[width=1.0\linewidth]{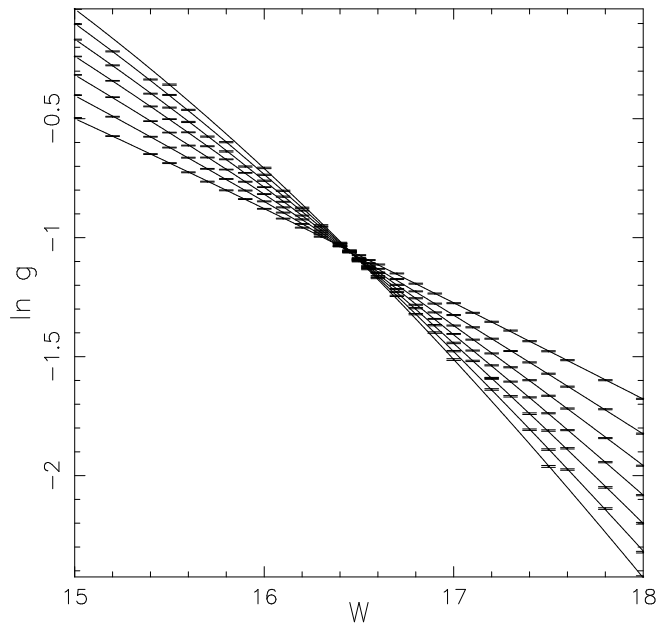}
\caption{Data for the median conductance ($q=0.5$ percentile), together 
with the best fit of Eq. \ref{fit}, as a function of disorder $W$ for 
systems sizes in the range $L$=6--18.}
\end{figure}

For the percentiles analysed the estimates of the critical disorder and 
the critical exponent
obtained from the scaling of different percentiles are consistent as required.
The estimates of the critical exponent in Table \ref{T1}
are consistent with our previous estimates
based on the scaling of the localisation length in quasi-$1d$ 
systems,\cite{slevin:99}
scaling of higher Lyapunov exponents,\cite{slevin:01b,markos:00} and
scaling of the mean conductance, mean resistance and typical 
conductance.\cite{slevin:01}
The estimates are also consistent with numerical estimates 
reported by other authors.\cite{mackinnon:94,romer:00,queiroz:00}

\begin{figure}[ht]
\includegraphics[width=1.0\linewidth]{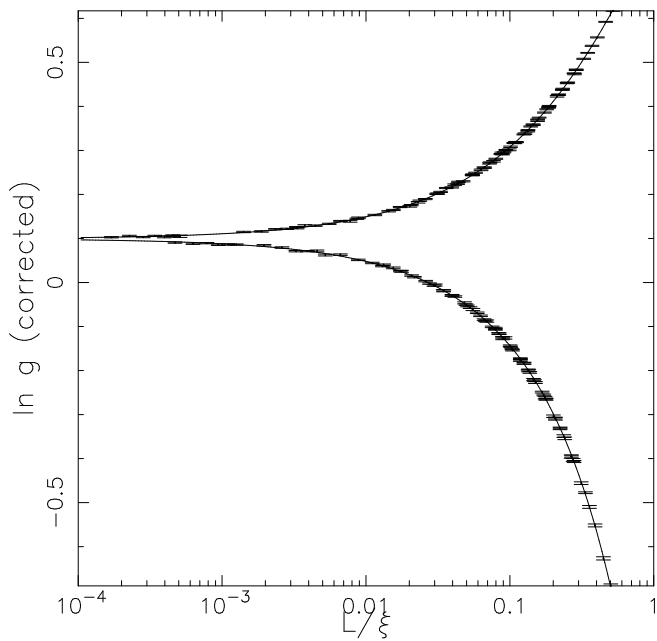}
\caption{The data for the $q=0.83$ percentile after corrections to scaling are subtracted and
plotted as function of the ratio $L/\xi$ to make single parameter scaling evident.
The lines are the scaling functions (\ref{fplus})and 
(\ref{fminus}) described in the text.}
\end{figure}

\section{Conclusion}

Our numerical results demonstrate single parameter scaling of the 
zero temperature conductance distribution in the critical regime 
of the Anderson transition in three dimensions.
This result complements a previous demonstration
of the scaling of the mean conductance, typical conductance and 
mean resistance.\cite{slevin:01}

A two parameter scaling of the conductance distribution, similar
to that found by by Deych {\it et al.} for one dimensional systems 
might be recovered in the strongly localised regime.
The localisation length diverges at the critical point while $l_s$,
which is related to the integrated density of states, is always finite.
Thus on approaching the critical point we should always find 
$\xi> l_s$ and single parameter scaling should be observed.
Far from the critical point, if $\xi$ becomes less than $l_s$, a two 
parameter scaling might appear. It remains to be seen, however, if
the results of Deych {\it et al.} carry over to higher dimensions.

\begin{acknowledgments}
We would like to thank the Institute for Solid State Physics 
of the University of Tokyo for the use of their computer facilities. 
PM would like to thank the Japan Society for the Promotion of
Science and Sophia University for their hospitality and financial
support, and is
also grateful for support under APVT Grant No. 51-021602.
\end{acknowledgments}

\bibliography{localisation}

\end{document}